\documentclass[a4paper,USenglish,cleveref, autoref, thm-restate]{lipics-v2021}
\hideLIPIcs  %uncomment to remove references to LIPIcs series (logo, DOI, ...), e.g. when preparing a pre-final version to be uploaded to arXiv or another public repository

\title{A Customized SAT-based Solver for Graph Coloring} %TODO Please add

\author{Timo Brand\footnote{Corresponding author}}{School of Computation, Information and Technology, Technical University of Munich, Germany}{Timo.Brand@tum.de}{https://orcid.org/0009-0004-3111-2045}{}
\author{Daniel Faber}{Department of Computer Science, University of Bonn, Germany}{dfaber@uni-bonn.de}{}%
{}%{This author was supported by the Deutsche Forschungsgemeinschaft (DFG, German Research Foundation) under grant FOR-5361 -- 459420781, and by the BMBF (Germany) and state of NRW as part of the Lamarr-Institute, LAMARR22B.}%authorspecific funding
\author{Stephan Held}{Research Institute for Discrete Mathematics and Hausdorff Center for Mathematics, University of Bonn, Germany}{held@dm.uni-bonn.de}{https://orcid.org/0000-0003-2188-1559}{}
\author{Petra Mutzel}{Department of Computer Science and Hausdorff Center for Mathematics, University of Bonn, Germany}{pmutzel@uni-bonn.de}{https://orcid.org/0000-0001-7621-971X}%
{}%{Same funding as Daniel Faber}%authorspecific funding

\authorrunning{T. Brand, D. Faber, S. Held, and P. Mutzel} %TODO mandatory. First: Use abbreviated first/middle names. Second (only in severe cases): Use first author plus 'et al.'

\Copyright{Timo Brand, Daniel Faber, Stephan Held, and Petra Mutzel} %TODO mandatory, please use full first names. LIPIcs license is "CC-BY";  http://creativecommons.org/licenses/by/3.0/

 %TODO mandatory: Please choose ACM 2012 classifications from https://dl.acm.org/ccs/ccs_flat.cfm

 %TODO mandatory; please add comma-separated list of keywords

\category{} %optional, e.g. invited paper

\relatedversion{To appear in the proceedings of \href{https://www.siam.org/conferences-events/siam-conferences/alenex26/}{ALENEX26}} %optional, e.g. full version hosted on arXiv, HAL, or other respository/website
\supplement{Source Code and evaluation scripts are available at: \url{https://doi.org/10.5281/zenodo.17328845}.}%TODO
\funding{\emph{Daniel Faber and Petra Mutzel}: Supported by the Deutsche Forschungsgemeinschaft (DFG, German Research Foundation) under grant FOR-5361 -- 459420781, and by the BMBF (Germany) and state of NRW as part of the Lamarr-Institute, LAMARR22B.}

\usepackage{mathtools}
\usepackage{amsmath}
\usepackage{csquotes}

\usepackage{tikz}
\usetikzlibrary{graphs, positioning} %graph drawing commands and relative node positioning
\usetikzlibrary{shapes.geometric} %allows for shapes such as regular polygons
\usepackage{booktabs}
\usepackage{supertabular}
\usepackage{float}
\usepackage[skip=2pt]{caption}
\renewcommand{\cref}{\Cref}

\newcommand{\lneg}[1]{\overline{#1}}
\DeclareMathOperator{\true}{True}
\DeclareMathOperator{\false}{False}

\DeclareMathOperator{\bag}{bag}
\DeclareMathOperator{\rep}{rep}

\def\copyrightline{}

\nolinenumbers
\EventEditors{}
\EventLongTitle{}
\EventShortTitle{}
\EventAcronym{}
\EventYear{}
\EventDate{}
\EventLocation{}
\EventLogo{}
\SeriesVolume{}
\ArticleNo{1}
\begin{document}

\maketitle

\begin{abstract}
We introduce ZykovColor, a novel SAT-based algorithm to solve the graph coloring problem
working on top of an encoding that mimics the Zykov tree.
Our method is based on an approach of H{\'e}brard and Katsirelos (2020) that employs a propagator
to enforce transitivity constraints, incorporate lower bounds for search tree pruning, and enable inferred propagations.

We leverage the recently introduced IPASIR-UP interface for CaDiCaL to implement these techniques with a SAT solver.
Furthermore, we propose new features that take advantage of the underlying SAT solver.
These include modifying the integrated decision strategy with vertex domination hints and
using incremental bottom-up search that allows to reuse learned clauses from previous calls.
Additionally, we integrate a more effective clique computation and an algorithm for
computing the fractional chromatic number to improve the lower bounds used for pruning during the search.

We validate the effectiveness of each new feature through an experimental analysis.
ZykovColor outperforms other state-of-the-art graph coloring implementations
on the DIMACS benchmark set.
Further experiments on random Erdős-Rényi graphs show that our new approach
matches or outperforms state-of-the-art SAT-based methods for both very sparse and highly dense graphs.
We give an additional configuration of ZykovColor that dominates other SAT-based methods on the Erdős-Rényi graphs.
\end{abstract}

\newpage
\section{Introduction}\label{sec:introduction}
Graph coloring consists of assigning a minimum number of colors to all
vertices of a graph $G$ such that no two adjacent vertices are assigned the same color.
The graph coloring problem is to find the chromatic number $\chi(G)$,
i.e., the smallest number of colors needed for such an assignment.
It is a classical combinatorial optimization problem with many applications~\cite{chaitin1982register,hadish2005jacobian,lofti1986schedueling}
that is notoriously difficult to solve.
It is NP-hard to approximate $\chi(G)$ within a factor $n^{1-\epsilon}$ for all $\epsilon > 0$~\cite{zuckerman2006linear}.

Various exact algorithms have been proposed to address the graph coloring problem.
Examples are the branch-and-bound variant DSATUR~\cite{brelaz1979new,san2012new,furini2017improved},
branch-and-price~\cite{mehrotra1996column,malaguti2011exact,gualandi2012exact,held2012maximum}
(based on the Zykov recursion~\cite{zykov1949some}),
integer programming~\cite{mendez2008cutting,cornaz2017solving,jabrayilov2018new},
SAT solvers~\cite{van2008another,glorian2019incremental,heule2022cliques,faber2024sat},
constraint programming solvers~\cite{hebrard2020constraint},
or binary decision diagrams~\cite{van2022graph}.

SAT solvers were employed successfully not only for exact coloring.
In recent work, a generic local search SAT solver found colorings matching known lower
bounds for the hard DIMACS instance wap07a~\cite{chowdhury2023linear}.

Recently, the IPASIR-UP interface~\cite{FazekasNPKSB24} for the SAT solver CaDiCaL~\cite{Cadical2024} was introduced.
It allows to modify the course of the SAT solver by registering callbacks to all major steps of the algorithm,
including custom propagation, adding clauses, decisions, and~lazy~reasoning.

Our approach is also based on the Zykov recursion~\cite{zykov1949some} that was already
used in conjunction with SAT solvers~\cite{schaafsma2009dynamic,glorian2019incremental}
and constrained programming~\cite{hebrard2020constraint}, where the authors introduce a hybrid
approach that combines constraint programming with SAT-solving techniques such as clause learning.

\subsection{Our Contribution}\label{subsec:our-contribution}

In this work, we adapt the propagator for the graph coloring problem presented in~\cite{hebrard2020constraint}.
Originally proposed in conjunction with a hybrid CP/SAT approach, we
adopt it to be used with the state-of-the-art SAT solver CaDiCaL~\cite{Cadical2024},
heavily using the recently introduced IPASIR-UP interface~\cite{FazekasNPKSB24} (\cref{subsec:transitivity-constraints,subsec:lower-bound-pruning,subsec:inferred-propagations}).
We use the features of the interface to present \textit{ZykovColor},
a customized SAT-based solver for the graph coloring problem.

Our new algorithms and their engineering include:
\begin{itemize}
    \item Implementing custom constraint propagations utilizing the IPASIR-UP interface with lazy (deferred) addition of reason clauses (\cref{subsec:transitivity-constraints}),
    \item using better clique heuristics for lower bound pruning (\cref{subsec:improved-cliques}),
    \item a new decision strategy, where we use vertex domination hints to modify the decisions made by the SAT solver (\cref{subsec:decision-strategy}),
    \item performing an incremental bottom-up search for the chromatic number  without re-initializing the solver for each decision problem
    (\cref{subsec:incremental-bottom-up-optimization}),
    \item using the fractional chromatic number to compute stronger lower bounds
        for more effective pruning on dense graphs (\cref{subsec:fractional-chromatic-number}).
\end{itemize}

We perform an ablation study demonstrating the effectiveness of each feature in ZykovColor (\cref{subsec:ablation-study}).
Additionally, we perform extensive computational studies, benchmarking ZykovColor with several state-of-the-art algorithms (\cref{subsec:evaluation-benchmark-graphs}).
On the DIMACS benchmarks for graph coloring, ZykovColor outperforms other state-of-the-art algorithms.
Additionally, we analyze the performance of different algorithms across varying graph densities using a large set of Erd\H{o}s-R\'enyi graphs.

\section{Preliminaries}\label{sec:preliminaries}

Let $G=(V,E)$ be an undirected graph with vertices $V$ and edges $E$.
In usual notation, $n\coloneqq |V|$,  $m\coloneqq |E|$,
and $N_G(u) = \{v \in V | \{u,v\}\in E \}$ is the set of neighbors of $u$,
shortened to $N(u)$ if the graph is clear from the context.
The (edge) density of  $G$ is the value $\frac{2m}{n(n-1)}$.

We assume that the vertices are numbered from 1 to $n$, i.e., \ $V =\{1,\dots,n\}$.
This allows us to give short precise descriptions of certain encodings later.
A coloring of size $k$ or $k$-coloring is a map
$f: V \to \{1,\ldots, k\}$ such that $f(u)\neq f(v)$ for all edges $\{u,v\}\in E$.
The graph coloring problem is to compute a coloring using the minimum possible number of colors;
the size of such a coloring is the chromatic number, denoted $\chi(G)$.

A clique in $G$ is a set of vertices such that each pair of vertices in the set is connected by an edge.
Thus, the size of any clique in $G$ provides a lower bound for $\chi(G)$.

To solve the graph coloring problem, we employ a satisfiability (SAT) solver
that takes a Boolean formula in conjunctive normal form as input.
It returns as the solution either SAT with a satisfying variable assignment or UNSAT if no satisfying assignment exists.
We will consider formulas that decide the $k$-coloring problem,
i.e., that are satisfiable if and only if a $k$-coloring exists.
Consequently, we must solve a series of decision problems to determine
the smallest $k$ such that the graph is $k$-colorable but not $(k-1)$-colorable.

This outer iteration might motivate the use of a MaxSAT solver instead.
However, we only consider SAT solvers for several reasons:
After determining lower and upper bounds in preprocessing, usually only a few decision problems need to be solved.
We also observed that for most instances, only the two problems $k \in \{\chi(G)-1, \chi(G)\}$ take a significant time.
Finally, we want to interact with the solving process.
For the SAT solver CaDiCaL~\cite{Cadical2024} we can use the IPASIR-UP interface~\cite{FazekasNPKSB24}.
We are not aware of similar interfaces in competitive MaxSAT solvers.

\subsection{Assignment Encoding}\label{subsec:assignment-encoding}

A natural satisfiability encoding to decide $k$-colorability is the assignment encoding.
It is called traditional encoding in~\cite{van2008another}.
It uses $n\cdot k$ variables $x_{vi} \in \{\true, \false\}$,
one for each vertex $v\in V$ and color $i \in \{1,\ldots, k\}$.
The variable $x_{vi}$ is $\true$ exactly if vertex $v$ is assigned color $i$.
The following clauses assert that a satisfiable assignment corresponds to a $k$-coloring,
while proving UNSAT shows that the graph has no $k$-coloring.
\begin{align}
    \bigvee_{i=1}^{k} x_{vi}           && v\in V \label{eq:assignment-encoding-1} \\
    \lneg{x}_{vi} \lor \lneg{x}_{vj}  && v\in V, 1\leq i < j \leq k \label{eq:assignment-encoding-3} \\
    \lneg{x}_{vi} \lor \lneg{x}_{wi}  && \{v,w\}\in E, i\in \{1, \ldots, k\} \label{eq:assignment-encoding-2}
\end{align}

\cref{eq:assignment-encoding-1,eq:assignment-encoding-3} ensure that each vertex is assigned exactly one color,
while \cref{eq:assignment-encoding-2} enforces that adjacent vertices get different colors.

This encoding is compact, but it contains a lot of symmetry.
Any permutation of the color classes in a $k$-coloring yields an equivalent coloring,
leading to a search space with $k!$ equivalent solutions.
To reduce these symmetries, symmetry-breaking constraints for the assignment encoding
were proposed in~\cite{heule2022cliques,faber2024sat} and in~\cite{mendez2008cutting}
in the context of integer programming.

\subsection{Zykov-based Encoding}\label{subsec:zykov-based-encoding}

\begin{figure}[tbp]
    \centering
    \includegraphics[width=0.45\linewidth]{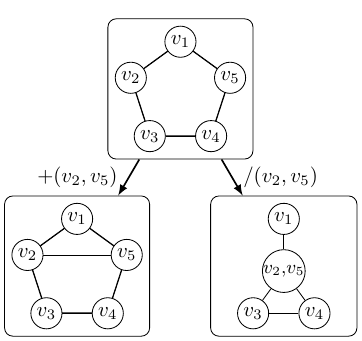}
    \caption{Root node and first two child nodes of the Zykov tree for the graph $C_5$.}
    \label{fig:zykov-tree-example}%
\end{figure}

An alternative encoding is based on the following Zykov recurrence for the chromatic number~\cite{zykov1949some}.
\begin{align}\label{eq:zykov}
    \chi(G) = \min\{\chi(G/(u,v)), \chi(G + (u,v))\} \; \text{ if } \{u,v\}\notin E
\end{align}
In an optimal coloring, two non-adjacent vertices $u,v$ will either have the same color or different ones.
These two cases are simulated by either merging the vertices ($G/(u,v)$) or by adding an edge between them ($G + (u,v)$).
An example of a Zykov recursion step for a graph is given in \cref{fig:zykov-tree-example}.
Recursive application yields a so-called Zykov tree, where all leaf vertices represent complete graphs.
Any leaf with $k$ vertices determines a $k$-coloring for the original graph
by assigning the same color to all vertices that were merged on the path from the root to the leaf in the Zykov tree.
Therefore, any leaf with a minimum number of vertices determines an optimal coloring and the chromatic number of the original graph.
The Zykov recursion was used successfully in branch-and-price frameworks~\cite{mehrotra1996column, malaguti2011exact,held2012maximum}.

In conjunction with a SAT solver it was first used in~\cite{schaafsma2009dynamic} to break symmetries
on top of the assignment encoding~\cref{eq:assignment-encoding-1,eq:assignment-encoding-2,eq:assignment-encoding-3}.
Later, the satisfiability encoding that we will discuss next became the basis of~\cite{glorian2019incremental} and~\cite{hebrard2020constraint}.

For each non-edge $\{u,v\}\notin E$, we define Boolean variables $e_{uv}$, where $e_{uv}= \true$
means that $u$ and $v$ are merged, and thus have the same color.
Likewise, $e_{uv}= \false$ means that the edge $\{u,v\}$ is added to the graph.
For ease of notation, we extend the variables $e_{uv}$ also to proper edges $\{u,v\} \in E$.
They are forced to $e_{uv} = \false$.

To ensure that each (partial) assignment of truth values corresponds to a valid graph in the Zykov tree,
transitivity constraints~\eqref{eq:transitivity} are required.
They assert that if $u,v$ and $v,w$ have the same color, vertices $u$ and $w$ must also have the same color.
With just the clauses from~\eqref{eq:transitivity},
any satisfying assignment corresponds to a leaf node in the Zykov tree
but the number of colors is not yet restricted.
To this end, Glorian et al. \cite{glorian2019incremental} use
auxiliary variables $c_v$ for each vertex $v\in V$.
In the  clauses~\eqref{eq:cv-def}, $c_v = \true$ means that vertex $v$ needs a new color
 different from any previous vertex in the order given by $V=\{1,\dots,n\}$.
Now, the sum over the $c_v$ variables counts the total number of used colors.
The \textit{at-most-$k$-constraints}~\eqref{eq:atmostk} can be modeled as SAT in multiple ways,
e.g., using cardinality networks~\cite{asin2011cardinality}
or the totalizer encoding~\cite{bailleux2003efficient}.
This yields the following \textit{full (Zykov) encoding} for the $k$-coloring problem.
\begin{align}
        \lneg{e}_{uv} \lor \lneg{e}_{vw} \lor e_{uw}       && u,v,w \in V\label{eq:transitivity}\\
        c_v \Leftrightarrow \bigwedge_{u < v} \lneg{e}_{uv} && v\in V \label{eq:cv-def}\\
        \sum_{v\in V} c_v \leq k \label{eq:atmostk}
\end{align}

A major drawback of this encoding is its size: it requires $O(\overline{E}) = O(n^2)$ variables
and $O(n^3)$ clauses just for the transitivity constraints~\eqref{eq:transitivity}.
Glorian et al.~\cite{glorian2019incremental} propose to start without transitivity clauses and iteratively
re-solve the SAT model, after adding transitivity clauses that were violated in the last solution.
H{\'e}brard and Katsirelos~\cite{hebrard2020constraint} instead propose a custom propagator
that adds the transitivity constraints on the fly while solving the SAT problem.
Their hybrid CP/SAT algorithm further combines clause learning and exploits the problem structure
to present an effective algorithm for the graph coloring problem.

\subsection{IPASIR-UP Interface}\label{subsec:ipasir-up-interface}

Before discussing the propagator for the graph coloring problem in detail,
we summarize the IPASIR-UP interface~\cite{FazekasNPKSB24} and important callbacks
that allow us to modify the solving process of the SAT solver CaDiCaL~\cite{Cadical2024}.

We briefly repeat the main steps in a Conflict-Driven Clause Learning (CDCL) solver:
\begin{itemize}
    \item \texttt{Boolean Constraint Propagation} and \texttt{Decide Literal} are repeated until either a satisfying assignment
            is found, or a clause becomes falsified under the current assignment.
    \item If a clause is currently falsified, \texttt{Conflict Analysis} produces a reason clause for the contradiction
            which is passed to \texttt{Clause Learning}.
    \item \texttt{Clause Learning}: If the derived clause is empty, it finishes in UNSAT.
            Otherwise, it is added and causes backtracking to a previous decision level and \texttt{Boolean Constraint Propagation} starts again.
\end{itemize}
When using an external propagator, any assignment found by the solver might violate some clauses from the user.
For this reason, CaDiCaL adds the artificial state \texttt{Solution Analysis} before transitioning to SAT,
where the external propagator can add additional clauses to the formula.
A more detailed description of CDCL can be found in~\cite{marques2021conflict}.

\begin{figure}[tbp]
    \centering
    \includegraphics[width=0.6\linewidth]{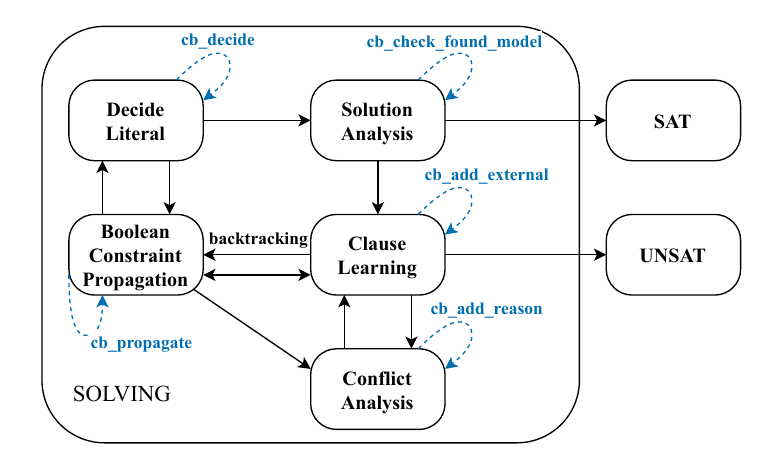}
    \caption{An overview of the states in the solving process and the provided user callbacks
        (Adapted from~\cite{FazekasNPKSB24}).}
    \label{fig:ipasir-up}
\end{figure}

The external propagator is notified by the SAT solver of newly assigned literals,
new decision levels, and backtracking.
It then has the possibility to modify the solving process using callbacks,
as visualized in \cref{fig:ipasir-up}.
For our application, we will use the following functionalities of the interface:
\begin{itemize}
    \item Provide literals that must be propagated (\texttt{cb\_propagate}),
    \item communicate which literal assignment is chosen as the next decision (\texttt{cb\_decide}), and
    \item add arbitrary clauses to the encoding (\texttt{cb\_add\_external}).
\end{itemize}
We need to supply a reason clause for any propagation, i.e.,
a clause such that all literals except the propagated one are false under the current assignment.
This is done with the \texttt{cb\_add\_reason} callback.
The IPASIR-UP interface allows us to add the reason clause lazily:
it is only required if the propagation is part of a conflict and needed in conflict analysis.

The remaining callback \texttt{cb\_check\_found\_model} is called when a satisfying assignment is found.
One can either accept the model or add further clauses to the problem.
While useful, we do not need this functionality in our approach, as the correctness of any satisfying assignment
will be ensured by the propagations and external clauses we add during the search.

\section{SAT-Based Propagator for the Zykov Encoding}
\label{sec:zykov-based-propagator}
In this section we adopt concepts of the propagator presented by H{\'e}brard and Katsirelos ~\cite{hebrard2020constraint}
to the satisfiability setting using the interface discussed in \cref{subsec:ipasir-up-interface}.
This includes details on the transitivity constraints,
using lower bounds for search tree pruning, and additional inferred propagations
(\cref{subsec:transitivity-constraints,subsec:lower-bound-pruning,subsec:inferred-propagations}).

We further extend the SAT implementation with novel ideas and improvements,
namely a better clique computation, modifications to the decision strategy,
the design of an incremental bottom-up optimization that can reuse clauses and learned information, and the computation of fractional chromatic numbers during search
(\cref{subsec:improved-cliques,subsec:decision-strategy,subsec:incremental-bottom-up-optimization,subsec:fractional-chromatic-number}).

\subsection{Transitivity Constraints}\label{subsec:transitivity-constraints}
We follow the approach in~\cite{hebrard2020constraint}
to avoid adding the cubic number of transitivity clauses~\eqref{eq:transitivity} of the full encoding.
Instead, the propagator ensures that $e_{uw}$ is set to $\true$ whenever $e_{uv}$ and $e_{vw}$ are set to $\true$.
This is done efficiently by maintaining a graph $H$ that corresponds to the graph $G^\prime$,
where $G^\prime$ is the graph in the Zykov tree obtained by following the merge operations and edge additions
given by the currently assigned variables $e_{uv}$.

Each vertex $v$ in $H$ stores a pointer to a bag/set $\bag(v)$ of merged vertices that it belongs to,
and the representative vertex $\rep(v)$ that it has been merged into, similar to a union-find data structure.
Initially, $\bag(v) = \{v\}$ and $\rep(v) = \{v\}$ for all vertices $v$ of $G$.

If the propagator receives the assignment $e_{uv} = \true$, $u$ and $v$ are merged and we propagate
$e_{u^\prime v^\prime} = \true$ for all $u^\prime \in \bag(u)$ and $v^\prime \in \bag(v)$.
Additionally, any vertex $u^\prime \in \bag(u)$ cannot have the same color as any of the vertices
$w \in N_H(\bag(v)) \backslash N_H(\bag(u))$ so that $e_{u^\prime w} = \false$ is also propagated,
and likewise for any $v^\prime \in \bag(v)$.
Finally, $\bag(u)$ and $\bag(v)$ are merged and the representative
for all merged vertices is updated to one of $\rep(u)$ or $\rep(v)$.

If the assignment is $e_{uv} = \false$, we simply add an edge between $u$ and $v$ to $H$ and
propagate $e_{u^\prime v^\prime} = \false$ for all $u^\prime \in \bag(u)$ and $v^\prime \in \bag(v)$.

These propagations are handled via \texttt{cb\_propagate},
and the explanation clauses~\eqref{eq:transitivity} are provided lazily using \texttt{cb\_add\_reason},
only should they be needed during conflict analysis.
This is an advantage over the CP/SAT hybrid in~\cite{hebrard2020constraint},
where transitivity clauses are immediately added to the encoding when the propagation is given.
An important observation is that the propagator considers each variable only once per search path.
This way, the transitivity propagation from the root node down to a full assignment only has complexity $O(n^2)$\cite{hebrard2020constraint}.
In contrast, the constraint propagation from the root to a full assignment
based on the full encoding of all transitivity constraints~\eqref{eq:transitivity} has cubic complexity.

\subsection{Lower Bound for Pruning}
\label{subsec:lower-bound-pruning}

Using a propagator to modify the solving process further allows us to dynamically prune the search tree.
Recall that during the search, the assigned variables $e_{uv}$ define a graph $G^\prime$,
which corresponds to a node in the Zykov tree. %this was mentioned in transitivity already
When a lower bound for the chromatic number of $G^\prime$ is found that is larger than $k$,
no coloring of size $k$ or smaller can be found by continuing the search on this graph.
The graph $G^\prime$ and its children can thus be pruned from the Zykov tree,
which means cutting off the current partial assignment and backtracking.

This lower bound-based pruning  eliminates all solutions with more than $k$ colors.
Thus, we do not need to add the clauses~\cref{eq:cv-def,eq:atmostk} explicitly
an can avoid the auxiliary $c_v$ variables.
If we find a lower bound that is larger than $k$, the assignment is pruned and the search is backtracked.

H{\'e}brard and Katsirelos~\cite{hebrard2020constraint} use cliques and a novel
Mycielski graph-based lower bound to determine when a node can be pruned from the search.
For cliques, they use a simple but fast heuristic.
It maintains a list of clique candidates and greedily adds vertices to any clique that allows it.
If the vertex cannot be added to any existing clique, a new singleton clique is created.
Finally, all vertices are re-examined and added to all newly added cliques that admit them.
In the original paper~\cite{hebrard2020constraint}, a vertex ordering to iterate over the vertices is computed in each call.
To avoid the resulting computational overhead, we instead use a fixed vertex ordering.
We put the clique computed in preprocessing at the beginning of the order
and fix the remaining vertices in the order in which they are selected by the DSatur heuristic.
By putting a clique first in the ordering, the heuristic always finds a clique at least as large,
which was not guaranteed by the dynamic vertex orderings considered in~\cite{hebrard2020constraint}.

For the second bound,~\cite{hebrard2020constraint} introduces a novel lower bound based on the Mycielskian of a graph.
Given a graph $G$, the Mycielskian of $G$ is a construction that has the same clique number as $G$
but chromatic number $\chi(G) + 1$.
The idea of~\cite{hebrard2020constraint} is to start with a subgraph $H$ of $G$ with chromatic number $k$
and try to identify the Mycielskian of $H$ as subgraph of $G$ which then has chromatic number $k+1$.
This construction can be applied iteratively and potentially provides significantly stronger bounds than a maximum clique.
H{\'e}brard and Katsirelos presented a greedy heuristic to find such generalized Mycielskians,
starting from a maximal clique as the initial subgraph.

To prune a partial assignment, we need to add a clause to the encoding
that is unsatisfied under the current assignment.
Since both the clique and Mycielskian bounds are based on a subgraph $H$,
the explaining clause can be given by a description of the responsible subgraph:
\begin{equation}
    \label{eq:bound-explanation}
    \bigvee_{\{u,v\} \in E(H)} e_{\rep(u)\rep(v)}
\end{equation}
It forces that at least one edge of $E(H)$ must be true, i.e. $H$ is destroyed.
The \texttt{cb\_add\_external} callback is used to add the clause to the encoding.
Because it is unsatisfied at the node where the subgraph is present and the lower bound is found,
this causes the desired backtracking.
The clique heuristic is called at every node of the search tree,
while the Mycielskian lower bound is only computed after backtracking
and if, the gap between clique and upper bound is at most one.
We did not observe a significant gain in pruning when the Mycielskian lower bound is computed regardless of the gap, in particular for bottom-up optimization.
We thus choose to call the heuristic less frequently.

\subsection{Inferred Propagations}
\label{subsec:inferred-propagations}
In this section, we discuss another strategy.
It does not prune the current partial assignment but infers necessary variable assignments.
Given a subgraph with chromatic number $k$ and vertices $u$ and $v$,
the propagator can determine that $e_{uv}$ must be assigned either $\true$ or $\false$,
if otherwise a subgraph with chromatic number $k+1$ is formed.

This is an application of two pre-processing rules that were, to our knowledge,
first presented in~\cite{lucet2004pre}.
They were re-discovered in~\cite{hebrard2020constraint} for use inside the propagator
and we adapt the two lemmas they stated for our propagator as well.

\begin{lemma}[Positive pruning, \cite{hebrard2020constraint}]
    \label{lem:pos-pruning}
    Given a subgraph $H = (V^\prime, E^\prime)$ in $G = (V,E)$ with chromatic number $k$,
    if there exists a vertex $v\in V \backslash V^\prime$ such that
    $V^\prime\backslash N(v) = \{u\}$ then $G + (u,v)$ has chromatic number at least $k+1$.
\end{lemma}
\begin{lemma}[Negative pruning, \cite{hebrard2020constraint}]
    Given a subgraph $H = (V^\prime, E^\prime)$ in $G = (V,E)$ with chromatic number $k$,
    if there exist two non-adjacent vertices $u$ and $v$ such that
    $V^\prime \subseteq (N(u) \cup N(v))$ then $G /(uv)$ has chromatic number at least $k+1$.
\end{lemma}

As before, the propagation is communicated using the \texttt{cb\_propagate} callback,
and the reason clause is based on the subgraph $H$, similar to~\eqref{eq:bound-explanation}.
The authors of~\cite{hebrard2020constraint} observed positive pruning to be cheap enough
to provide a benefit while negative pruning was too expensive.
We confirmed this in initial experiments with our implementation and thus only use the former.
Furthermore, we only choose cliques as subgraphs for pruning and not Mycielskians.

\subsection{Improved Clique Computation}
\label{subsec:improved-cliques}
While very fast, the clique heuristic used in~\cite{hebrard2020constraint} (see \cref{subsec:lower-bound-pruning})
is a greedy algorithm.
It often does not produce near-optimum cliques.
To improve the pruning potential of the propagator, we prefer an algorithm that produces larger cliques.
Since the clique algorithm is called frequently in the propagator, keeping its running time small is essential.

Local search algorithms have proven to be effective in finding larger cliques~\cite{galinier2006survey}.
We use the multi-neighborhood tabu search (MNTS) algorithm from~\cite{wu2012multi},
which combines local search with tabu search.
The algorithm starts from a randomly generated clique.
The local search consists of three operations, dropping a vertex from the clique, adding a vertex to the clique,
and replacing a vertex in the clique by another vertex.
We ported their source code\footnote{\url{https://leria-info.univ-angers.fr/~jinkao.hao/clique.html}}
from C to C++ and modified it to work with our data structures.
We choose deterministically changing seeds for each call to MNTS to obtain a deterministic algorithm.

Their algorithm has two parameters, the maximum number of iterations $\text{Iter}_{\max}$ for the local search and the search depth $L$ before restarting.
In their original experiments for the unweighted maximum clique problem, they were set to $\text{Iter}_{\max}=10^8$ and $L=10^4$.
Since we have frequent  invocations, we prioritize lower running times and chose
 $\text{Iter}_{\max} = 200$ and $L = 25$, providing a good balance as determined in initial experiments.

Inside the propagator, we first use the fast greedy algorithm.
MNTS is called only if the greedy algorithm does not find a clique that allows pruning.

\subsection{Improved Decision Strategy}
\label{subsec:decision-strategy}
H{\'e}brard and Katsirelos~\cite{hebrard2020constraint} further present a problem-specific decision strategy
that emulates the well-known DSatur heuristic for graph coloring.
In our experiments with CaDiCaL~\cite{Cadical2024}, we observed a better performance when not using such a custom decision strategy.
We stick with CaDiCaL's default decision strategy except for the following modification.

We  use the reduction rule of dominated vertices to
determine \enquote{good} decisions and communicate them to the SAT solver.
A vertex $u$ is dominated by a vertex $v$ if $N(u) \subseteq N(v)$
and we know that vertex $u$ can always take the same color as the more constrained vertex $v$.
We can therefore merge $u$ into $v$ as there exists an optimal solution
where the two vertices have the same color.
Note that we can potentially also find an optimal solution where $u$ and $v$ have different colors.
Thus, domination is weaker than the inferred propagations from the previous section.
Therefore, we provide $e_{uv} = \true$ as a decision for the
SAT solver using the \texttt{cb\_decide} callback.
The idea is that following this decision will lead to a part in
the search tree that is more likely to contain a solution and thus should be checked first.

Checking every pair of vertices for domination would require $O(n^2)$ running time.
To speed this up, we track the vertices whose neighborhood has grown since the last decision was made
and only check if one of those vertices now dominates a new vertex.

\subsection{Incremental Bottom-Up Optimization}
\label{subsec:incremental-bottom-up-optimization}
We previously highlighted that our implementation using the IPASIR-UP interface for
the Zykov propagator is satisfiability-based contrary to the original hybrid CSP/SAT approach.
It also allows us to use SAT solver techniques such as assumptions for incremental calls to the solver.
Assumptions allow us to fix a variable to a specific value throughout a call to the solver.
In the next call, the assumptions are reset, and the variable is allowed to take any value again.
This can be used to activate and deactivate constraints between calls.

In particular, we make iterative calls to the SAT solver with increasing guesses $k$ of the chromatic number.
SAT solvers do not allow the removal of clauses and thus constraints, but clauses can be deactivated through assumptions.
The clause of \cref{eq:bound-explanation} that is used
to prune subproblems can be extended with an activation literal $s_k$ as follows.
\begin{equation}
    \label{eq:bound-with-assumption}
    \left(\bigvee_{\{u,v\} \in E(H)} e_{\rep(u)\rep(v)}\right) \lor s_k
\end{equation}
When looking for a $k$-coloring, the literal $s_k$ is assumed to be $\false$ so that
\cref{eq:bound-with-assumption} will simplify to \cref{eq:bound-explanation} and can be used for pruning as normal.
If we are not looking for a $k$-coloring anymore,
we want to disable the clause and we do so by adding the unit clause $(s_k)$.
The clause of \cref{eq:bound-with-assumption} is now satisfied and will not cause any incorrect pruning.

Incrementally changing the constraint for the number of colors instead of re-initializing the SAT solver from scratch,
allows us to reuse the already added transitivity clauses and conflict clauses.
The SAT solver can keep useful information such as the variable activity, which is used in the decision strategy.

\subsection{Lower Bounds Using the Fractional Chromatic Number}
\label{subsec:fractional-chromatic-number}

The clique lower bound is not particularly strong, e.g., Mycielski graphs have clique number 2 and arbitrary
high chromatic numbers.
A stronger bound is the fractional chromatic number of a graph with a gap of $\mathcal{O}(\log n)$~\cite{lovasz1975ratio}.

The fractional chromatic number $\chi_f(G)$ is the value of the linear relaxation of the following stable set cover formulation,
where $\mathcal{S}$ denotes the set of all stable sets in $G$.
\begin{align}
\chi(G) = & \min  && \sum_{S\in \mathcal{S}}x_S\label{eq:stable-set1}   &&\\
& \text{s.t.} && \sum_{S\in \mathcal{S}: v \in S} x_{S} \geq 1   &&  \forall v \in V\label{eq:stable-set2}  &\text{(CIP)}\\
&             && x_S \in \{0,1\}  &&  \forall S \in \mathcal{S} \label{eq:stable-set3},
\end{align}
Clearly, $\lceil\chi_f(G)\rceil \le \chi(G)$ provides a lower bound.

As a consequence of~\cite{zuckerman2006linear}, it is almost as hard to approximate as the chromatic number itself.
However, in practice, it is often tractable using column generation~\cite{mehrotra1996column,held2012maximum,malaguti2011exact}.

Unlike the clique and Mycielskian bounds, this bound does not yield a specific subgraph that needs to be avoided.
Thus, we simply add the trivial clause containing the negation of every decision in the current node to trigger the desired backtracking.
This clause only prunes the current node and will not be useful in other parts of the search tree.

We use exactcolors~\cite{held2012maximum} to compute the fractional chromatic number bound $\lceil\chi_f(G)\rceil$.
Despite using a linear programming solver, it guarantees numerically safe lower bounds like all other methods used in our work.
Computing the fractional chromatic number can be time-consuming. Usually, it is  more efficient on dense graphs.
Therefore, we only compute this bound after backtracking and if the density of the graph $G'$ in the current Zykov node
is greater or equal to $0.75$.
The threshold was validated experimentally for the best performance profile on DIMACS benchmarks.
On Erd\H{o}s-R\'enyi graphs lower thresholds can be favorable as we will demonstrate in Section~\ref{subsec:evaluation-benchmark-graphs}.

\section{Experimental Evaluation}\label{sec:experimental-evaluation}

In our experimental evaluation, we are interested in answering the following questions:
\begin{itemize}
    \item Q1: What is the performance gain of the SAT-based algorithm and each new feature? %ablation study
    \item Q2: How does the performance of our new algorithm ZykovColor compare with state-of-the-art graph coloring algorithms?
    What are the strengths of the various algorithms?
\end{itemize}

To answer Q1, we perform an ablation study with different configurations of ZykovColor.
For Q2, we compare ZykovColor and several other algorithms on the DIMACS benchmark set of instances
and a large set of generated Erd\H{o}s-R\'enyi graphs with different densities.

\subsection{Implementation Details of ZykovColor}\label{subsec:implementation}

\subparagraph{Graph reduction techniques}
Before passing the graph to the exact coloring algorithm,
we apply reduction techniques to obtain a smaller graph $G^\prime$
with fewer vertices and edges and the same chromatic number.
The following two steps are commonly used~\cite{lucet2004pre,hebrard2020constraint,faber2024sat}
and iteratively applied to the graph until no more vertices can be removed.
\begin{itemize}
    \item \textbf{Low-degree Vertices}
    If $k$ is a lower bound on the chromatic number and
    $v$ is a vertex of degree less than $k$, we can remove $v$ from $G$.
    When recovering this reduction, $v$ can be assigned one of the colors not used by any of its neighbors.
    \item \textbf{Dominated Vertices}
    A vertex $u$ is dominated by another vertex $v$ if $N(u) \subseteq N(v)$.
    The dominated vertex can be removed from $G$ and later be assigned the same color as $v$
    since it is adjacent to the same or more vertices than $u$.
\end{itemize}

Note that dominated vertices might occur again later during the Zykov recursion.
Then, merging will be favored as a decision as described in \cref{subsec:decision-strategy}.

\subparagraph{Initial upper and lower bounds}
Our implementation uses the exact maximum clique algorithm CliSAT\footnote{\url{https://github.com/psanse/CliSAT}}~\cite{san2023clisat}
with a time limit of one second to compute a large initial clique,
and an initial call to the Mycielskian lower bound procedure to compute a strong lower bound $lb$.
The Dsatur coloring heuristic is used for a quick upper bound $ub$~\cite{brelaz1979new}.
If the instance is not yet solved, the lower bounds are used in the iterative application of the reduction rules,
after which the lower and upper bound computations are called again if any vertices were removed.
Additionally, exactcolors is called to compute the fractional chromatic number
on the reduced graph with a time limit of 10 seconds.
Should exactcolors not be able to report $\chi_f(G)$ within the 10 seconds,
it is deemed too expensive to run inside the propagator,
and the bounding technique described in \cref{subsec:fractional-chromatic-number} is disabled.
Finally, if this does not already solve the instance, the reduced graph is passed to the SAT solver.

\subparagraph{Optimization strategy}
We determine the chromatic number using bottom-up search:
We solve the $k$-coloring problem for $k = lb, lb+1, \ldots, ub - 1$
until the first satisfiable problem is found.
Performing bottom-up search has an advantage over top-down or binary search for our pruning-based algorithm
since it allows us to always use the tightest possible upper bound $k$.
The advantage of this is also confirmed experimentally in \cref{subsec:ablation-study}.

\subparagraph{Solving the Decision Problem}
For solving the $k$-coloring problem, we use the propagator
as described in \cref{sec:zykov-based-propagator}.
In particular, it handles the transitivity constraints and lazily adds the necessary clauses.
It uses the clique and Mycielskian lower bounds to dynamically prune the search;
both a greedy heuristic and local search are invoked to find large cliques.
Further, positive pruning using cliques as subgraph is enabled,
see \cref{lem:pos-pruning}.
The fractional chromatic number is computed as lower bound in certain nodes, see \cref{subsec:fractional-chromatic-number}.
Instead of a custom decision strategy, we use the default strategy of CaDiCaL,
with the addition of deciding on dominated vertices first, if available.
The SAT feature of assumptions is used to enable incremental bottom-up solving,
which allows the solver to reuse information in the next decision problem.
Finally, we use version 2.1.2 of CaDiCaL as the underlying satisfiability solver.
We refer to our algorithm in the configuration as described above as \textbf{ZykovColor} (short name \textbf{ZC}).

\subsection{Test Setup}\label{subsec:test-setup}

All experiments were run on an Intel Xeon Platinum 8480+ Sapphire Rapids with 512 GB of memory
(Benchmarks~\cite{Benchmark} user time: r500.5=3.35s),
and all sources were compiled using gcc 14 and the -O3 optimization flag, or run with Python3.9.
Unless stated otherwise, all experiments were run with a time limit of one hour and no memory limit.

\subparagraph{Benchmark instances}
The DIMACS benchmark set~\cite{johnson1996cliques} is the standard benchmark set for graph coloring.
It comprises 137 instances, which include real-world graphs, graphs with a certain structure, and random graphs.
We use it as the main benchmark set.

In addition, we conduct experiments on random Erd\H{o}s-R\'enyi graphs~\cite{erdds1959random}.
Given a natural number $n$ and a probability $p\in [0,1]$, an Erd\H{o}s-R\'enyi graph $G(n,p)$
is constructed by taking $n$ vertices and connecting any two vertices with independent probability $p$.
These graphs have an expected density of $p$ and were also considered in~\cite{jabrayilov2018new,hebrard2020constraint,san2012new}.
In our experiments, we consider 100 graphs $G(n,p)$ for every combination
of the following parameters, totaling 5000 graphs.
\begin{align*}
    n \in &\{70, 80, 90, 100, 110\}\\
    p \in &\{0.05, 0.1, 0.15, 0.2, 0.25, 0.3, 0.4, 0.5, 0.7, 0.9\}
\end{align*}
We chose slightly more sparse parameter configurations, reflecting the density distribution of many graphs in the
DIMACS benchmark set and in real-world applications.
Further, the numbers of vertices were chosen to produce reasonably difficult graphs.
The script to generate these graphs can be found with the source code of ZykovColor.

\subparagraph{State-of-the-art Algorithms}
For comparison, we chose recent exact solvers that achieved strong results and have published their source code.
We ran each algorithm on the same machine to get comparable running times.

The first choice is the original implementation of the Zykov propagator \textbf{gc-cdcl}\footnote{\url{https://bitbucket.org/gkatsi/gc-cdcl/src/master/}}~\cite{hebrard2020constraint},
which uses the constraint programming solver MiniCSP\footnote{\url{https://bitbucket.org/gkatsi/minicsp}}.
We also compare with two other satisfiability algorithms:
\textbf{POP-S}\footnote{\url{https://github.com/s6dafabe/popsatgcpbcp}}~\cite{faber2024sat},
using Kissat 3.1.1\footnote{Replacing the solver with version 4 of Kissat or Version 2.1.2 of CaDiCaL resulted in fewer solved instances}%
\cite{kissat2022solvers} to solve a partial order encoding of the problem,
and \textbf{CliColCom}\footnote{\url{https://github.com/marijnheule/clicolcom}} (short name \textbf{CCC})~\cite{heule2022cliques},
which alternates between computing larger cliques and colorings with fewer colors.
To find improved colorings or prove optimality,
they use either a local search solver or CaDiCaL 1.5.2%
\footnote{Replacing the solver with version 2.1.2 of CaDiCaL resulted in errors for some instances and ultimately solved fewer instances}
on the assignment encoding.

Further, we also implemented the assignment encoding from (\ref{eq:assignment-encoding-1})--(\ref{eq:assignment-encoding-3}).
It is embedded in the same framework as ZykovColor,
using the same SAT solver CaDiCaL, initial lower and upper-bound heuristics, initial reductions,
and bottom-up search strategy.
We add the relatively weak symmetry-breaking constraints~\eqref{eq:weak-symmetry}
and fix the color of the vertices in the clique $C$ found in preprocessing to exclude some symmetric solutions.
This reduces the number of symmetries to $(k-|C|)!$.
\begin{align} \label{eq:weak-symmetry}
    \lneg{x}_{vi} && \forall v < i, i \in \{2, \ldots, k\}
\end{align}
We refer to this algorithm as \textbf{Assignment} (\textbf{Ass}).

We also include the branch-and-price implementation \textbf{exactcolors}\footnote{\url{https://github.com/heldstephan/exactcolors}} (\textbf{EC})~\cite{held2012maximum},
using Gurobi 12.0.0 as the underlying linear programming solver.
Note that inside ZykovColor, we use exactcolors without branching, i.e.,
only to compute the fractional chromatic number.
In the comparison here, it uses branching to compute the chromatic number.
One of the oldest paradigms for exact graph coloring is the branch-and-bound algorithm
\textbf{DSatur}~\cite{brelaz1979new,san2012new}.
We use the improved version\footnote{\url{https://bitbucket.org/gkatsi/gc-cdcl/src/master/sota/Segundo/DSATUR/}}
suggested by~\cite{san2012new}.

\textbf{Picasso}~\cite{glorian2019incremental} would be another SAT-based algorithm
that uses the Zykov encoding and claims strong results.
Unfortunately, the published source code is \enquote{currently incorrect}~\cite{glorian2019incremental_sourcecode}.
Our attempt to implement it performed significantly worse.
Thus, we do not include it in our experiments.

\subsection{Ablation Study}\label{subsec:ablation-study}

We proposed several new features to obtain the ZykovColor algorithm
and now want to analyze their individual contributions.
Additionally, we want to evaluate the impact of the search tree pruning.
To this end, we consider ZykovColor and the following configurations each having one feature changed or disabled,
and the last two variants that use no pruning at all.
\begin{itemize}
    \item \textbf{No MNTS}: disabling the MNTS clique heuristic (described in \cref{subsec:improved-cliques})
    \item \textbf{Clique decision}: Instead of the decision strategy described in \cref{subsec:decision-strategy},
            use the strategy of gc-cdcl (Subsection 4.4 of~\cite{hebrard2020constraint}) based on large cliques.
            It is motivated by DSatur's branching heuristic.
    \item \textbf{No dominated}: not using dominated vertices as decisions (\cref{subsec:decision-strategy}), CaDiCaL is making all decisions.
    \item \textbf{Non-incremental}: the solver is re-initialized for each $k$-coloring problem.
    \item \textbf{Top-down}: search is performed top-down instead of bottom-up.
    \item \textbf{No-Fractional}: disable the computation of the fractional chromatic number
            using exactcolors (\cref{subsec:fractional-chromatic-number})
    \item \textbf{Full Encoding}: encoding with all clauses~\eqref{eq:transitivity}-\eqref{eq:atmostk} added,
        using the totalizer encoding~\cite{bailleux2003efficient} of Open-WBO\footnote{\url{https://github.com/sat-group/open-wbo}}~\cite{martins2014open}
        for the at-most-$k$ constraints. No lower bound pruning is applied.
    \item \textbf{Transitivity only}: same as the Full Encoding,
        but instead of adding all transitivity clauses explicitly, they are added and handled by the propagator.
\end{itemize}

\begin{figure}[tbp]
    \centering
    \includegraphics[width=\linewidth]{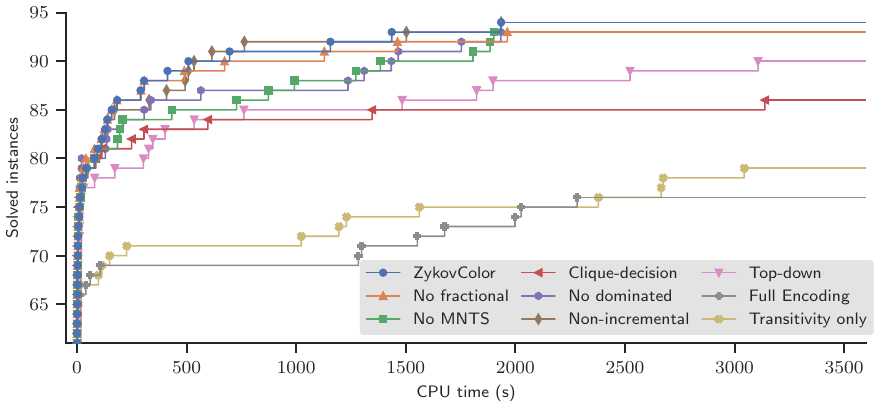}
    \caption{Survival plot of ablation study for ZykovColor variants and Full Encoding/Transitivity only.}
    \label{fig:survival-ablation}
\end{figure}

\cref{fig:survival-ablation} provides a survival plot for the performance of
the considered configurations on the DIMACS benchmark graphs.
The $x$-axis shows the time limit in seconds,
and the $y$-axis shows the number of solved instances within that time limit.
Each marker represents an instance that was solved at the time of the $x$-coordinate.
This allows one to compare the performance of different algorithms
for any time limit of at most one hour and gives a quick overview of their performance profile.

We can see that pruning of the search tree during solving is the most essential feature of the propagator approach,
as both the Full Encoding and the propagator that only handles the transitivity clauses solve far fewer instances.
This shows the capabilities of employing the IPASIR-UP interface.
Instead of treating the SAT solver as a black box, we can significantly improve the performance
by communicating information that is available to the user but not to the SAT solver.
Between the Full Encoding and Transitivity only,
we observe that even just handling the transitivity constraints with a propagator
has an advantage over the full satisfiability encoding.

Considering the ZykovColor variants, we observe the configurations
Top-down and Clique-decisions to solve fewer instances than ZykovColor in one hour,
with 89 and 84 instances solved, respectively.
The decision strategy appears to play a significant role;
despite the clique-based decision strategy being designed for the graph coloring problem,
the general SAT solver strategy performs much better here.
Regarding Top-down, it leads to less pruning
of the search space since a larger lower bound is required for a conflict,
and the performance decreases noticeably.

The variants No Fractional, No MNTS, and No dominated perform similarly but slightly worse than
ZykovColor and Non-incremental, solving 93 instances overall.
The latter two configurations perform very similarly as well, and both solve 94 instances
(the last instance is solved at the same time).
ZykovColor shows slightly better performance for lower time limits
and manages to solve some instances faster than Non-incremental.
Thus, we choose ZykovColor as the default configuration for our implementation.

\subsection{Comparison with State-Of-The-Art Coloring Algorithms}
\label{subsec:evaluation-benchmark-graphs}
\subparagraph{DIMACS Benchmarks}

\begin{figure}[tbp]
    \centering
    \includegraphics[width=\linewidth]{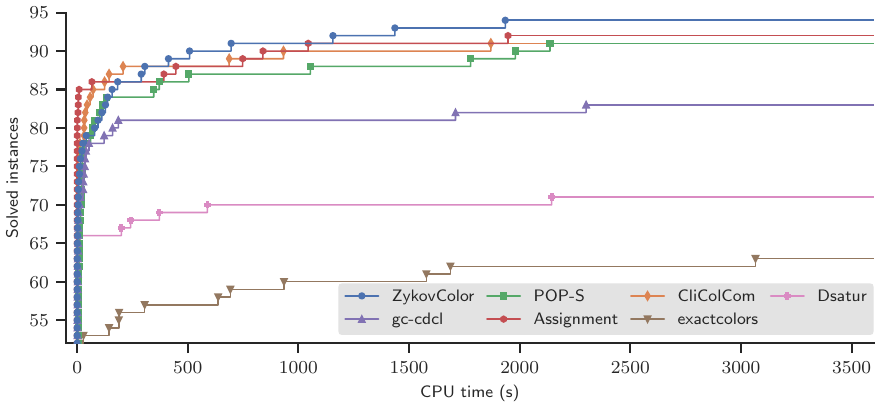}
    \caption{Survival plot for different algorithms on the DIMACS graphs.}
    \label{fig:dimacs-survival}
\end{figure}

\cref{tab:dimacs-overview} lists the number of solved instances out of the 137 graphs
from the DIMACS benchmark set for each considered algorithm.
A detailed table that lists the running time for each solved instance can be found in \cref{tab:dimacs-data} in the appendix.
We visualize the results with a survival plot in \cref{fig:dimacs-survival},
which allows us to compare the performance for different time limits.

We can see that the SAT-based methods perform the best for any time limit,
followed by the CSP/SAT hybrid gc-cdcl,
the branch-and-bound method DSatur and the branch-and-price algorithm exactcolors.

ZykovColor, the algorithm based on the improvements presented in this paper,
performs significantly better than gc-cdcl.
This shows the positive impact of the new features presented in this work.
Further, it solves three more instances than POP-S or CliColCom, the best-performing methods from the literature,
and two more instances than the implementation of the assignment encoding.
Assignment and CliColCom are more successful than ZykovColor only for small time limits.
ZykovColor outperforms CliColCom for time limits from 400 seconds and Assignment for time limits from 300 seconds.
Ultimately, all SAT methods perform similarly well, each solving between 91 and 94 instances.

The encoding plays an important role, but differences in the preprocessing,
the implementation, and the used solver also affect the performance.
Notably, the Mycielskian bound used in the preprocessing of ZykovColor and Assignment
can solve the difficult instance \texttt{Myciel7}, which POP-S and CliColCom cannot solve.

\begin{table}
    \setlength{\tabcolsep}{2.75pt}
    \centering
    \caption{Number of solved DIMACS instances within one hour.}
    \label{tab:dimacs-overview}
    \begin{tabular}{@{\extracolsep{\fill}}c|ccccccc@{}}
        {DIMACS} & {ZC} & {gc-cdcl} & {POP-S} & {Ass} & {CCC} & {EC} & {DSatur}\\
        \midrule
        137 & 94 & 83 & 91 & 92 & 91 & 63 & 71\\
    \end{tabular}
\end{table}

\subparagraph{Erd\H{o}s-R\'enyi Graphs}

\begin{figure}[btp]
    \centering
    \includegraphics[width=\linewidth]{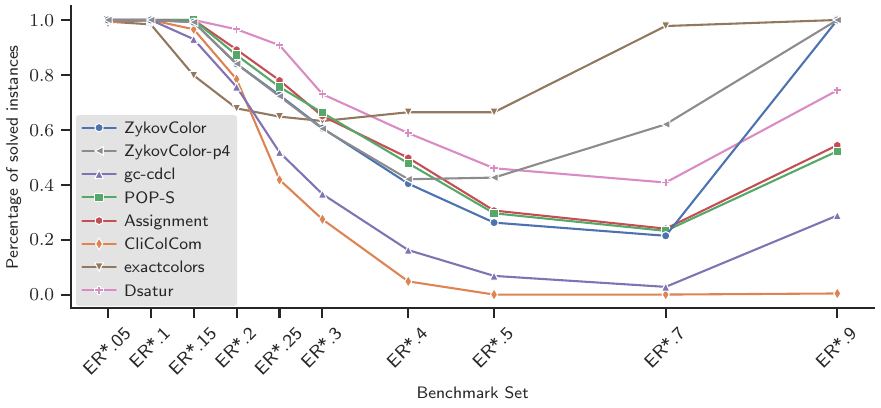}
    \caption{Normalized algorithm performance by parameter for ER graphs.}
    \label{fig:er-performance}
\end{figure}

We now compare the performance of different methods on the Erd\H{o}s-R\'enyi graphs.
We are particularly interested in density-dependent performances.
To visualize this, we use a parallel coordinate plot in \cref{fig:er-performance}.
The set of instances is given on the $x$-axis, where $ER^*.p$ represents
all Erd\H{o}s-R\'enyi graphs of any size for the edge probability $p$.
The $y$-axis represents the ratio of solved instances within the time limit of one hour.

We can see that all algorithms except for exactcolors exhibit similar performance curves depending on the density parameter $p$.
Initially, the performance decreases with increasing density until $ER^*.7$, after which the
performance again increases for the very dense instances $ER^*.9$.
Exactcolors follows a different trend.
The performance is roughly constant on graphs $ER^*.2$ to $ER^*.5$.
Then, it increases for $ER^*.7$ and $ER^*.9$.
The sparsest instances $ER^*.05$ and $ER^*.1$ are easy for all algorithms.

Looking at individual algorithm performance, we notice that DSatur and exactcolors,
the worst two algorithms on the DIMACS graphs, are now performing much better.
DSatur outperforms the SAT-based methods and gc-cdcl on all sets,
and exactcolors becomes the best method for graphs of density $0.4$ and higher.

Assignment, POP-S, and ZykovColor perform similarly.
Assignment is slightly better, and ZykovColor is slightly behind the two other methods.
However, ZykovColor shows an advantage on very dense graphs.
For the $ER^*.9$ graphs,
the fractional chromatic number is used as bound after every backtrack
(as the density of the graphs is always above the threshold of 0.75),
which notably increases the performance compared to Assignment and POP-S.

CliColCom performs significantly differently compared to on the DIMACS graphs.
It only shows good performance for the sparsest graphs up to $ER^*.2$.

As exactcolors performs very well on the Erd\H{o}s-R\'enyi graphs,
we attempt to leverage its strength by computing the
fractional chromatic number inside ZykovColor more frequently.
We do so by reducing the density threshold from $0.75$ to $0.4$,
as exactcolors becomes the best algorithm at $ER^*.4$.
This configuration is labeled ZykovColor-p4 in \cref{fig:er-performance} and
we can see that this considerably improves the ratio of solved instances for $ER^*.4$ to $ER^*.7$.

\subparagraph{Impact of the Threshold on Erd\H{o}s-R\'enyi Graphs}
\label{subsec:threshold-discussion}

Using the fractional chromatic number lower bound with a smaller density threshold
makes ZykovColor more competitive on Erd\H{o}s-R\'enyi graphs than other SAT approaches.
It is unclear how one would leverage this bound in the assignment or other related SAT formulations.

We did not see a benefit in using a lower density threshold on the DIMACS instances.
The running time spent on computing fractional lower bounds rather lowers the performance profile.

However, on Erd\H{o}s-R\'enyi graphs exactcolors remains the leading algorithm.
We conjecture that one reason is the lack of structure in the random graphs.
Here, the learned conflict clauses are likely less useful,
and a strong bound such as the fractional chromatic number might be more decisive.

Another aspect could be the branching strategy.
In exactcolors, the fractional covering solution is used to make branching decisions,
which shows good results for the random Erd\H{o}s-R\'enyi graphs.
Modern SAT solvers use activity scores of the variables for their branching decision,
which are influenced by the learned clauses.
Again, if the instance is random and the learned clauses are less meaningful,
the decisions made in the SAT solver might not be as good as those made by exactcolors.
On the instance DSJC125.9, exactcolors visits 114 Zykov nodes, while ZykovColor visits 7400 nodes.
Using the fractional chromatic number bound in every Zykov tree node and not only after backtracking
would reduce this to 1000 nodes.

\subparagraph{Memory Usage}
\label{subsec:memory-usage}
On the DIMACS graphs, ZykovColor, Assignment, and gc-cdcl
used less than 4.5, 7.5, and 8.5 GB, respectively,
except for the instance \texttt{qg.order100}, where the consumption was 25, 25, and 44 GB.
The average consumption was  385, 685, and 825 MB.

It is surprising that Assignment uses more memory than ZykovColor, despite the compactness of the encoding.
The Zykov-based encoding needs $O(n^2)$ variables, one for each non-edge of the graph. This could become significant for large sparse graphs.
Despite many such graphs in the benchmark set, ZykovColor requires noticeably less memory.
This indicates a more effective traversal of the search tree and potentially fewer or shorter added conflict clauses.

For the small sizes of the Erd\H{o}s-R\'enyi graphs, the algorithms ZykovColor, Assignment, and gc-cdcl all use about 60 MB of memory on average.

\section{Conclusion}\label{sec:conclusion}

We presented \textbf{ZykovColor}, a new graph coloring algorithm
using a SAT encoding that simulates the Zykov recurrence.
We adopted the propagator presented in~\cite{hebrard2020constraint} and enhanced it with
lazily added reason clauses using the IPASIR-UP interface~\cite{FazekasNPKSB24} of CaDiCaL~\cite{Cadical2024}.
We also proposed a modified decision strategy of the underlying SAT solver CaDiCaL~\cite{Cadical2024}.
A decisive feature is the use of stronger lower bounds for pruning.
For assignment or partial order-based SAT encodings, it would be unclear how to leverage them.

In an ablation study we demonstrated the impact of the individual ideas.
The biggest improvement arises from using a propagator for pruning.
Additionally, the decision strategy has a significant impact on algorithm performance.
Here, the (modified) strategy of CaDiCaL performs better than the problem-specific strategy from~\cite{hebrard2020constraint}.
Further, bottom-up optimization solves several more instances than top-down optimization,
contrary to the results of gc-cdcl~\cite{hebrard2020constraint}, where both configurations performed similarly.

ZykovColor outperforms other exact graph coloring approaches from the literature on the DIMACS benchmark set.
On dense (random) Erdős-Rényi graphs, SAT-based methods still fall behind  branch-and-price.
This may motivate possible directions for further improving SAT-based methods.

\bibliography{bibliography}

\appendix

\section{Detailed Results on DIMACS Instances}\label{sec:detailed-results-on-dimacs-instances}
In this appendix, we provide detailed results on the individual DIMACS instances (Table~\ref{tab:dimacs-data}).
{%input table and change spacing
    \renewcommand*{\arraystretch}{0.95}
	\scriptsize
	\setlength{\tabcolsep}{2.75pt}
    \begin{center}
\tablecaption{
    Instance data (number of vertices $n$, number of edges $m$ and density $d$)
    and running time of algorithms for solved DIMACS instances.
    The best time for each instance is highlighted in bold.
}
\tablehead{
\toprule
Instance & $n$ & $m$ & $d$ & ZykovColor & gc-cdcl & POP-S & Assignment & CliColCom & exactcolors & Dsatur \\
\midrule
}
\tabletail{
\midrule
\multicolumn{8}{r}{Continued on next page} \\
\midrule
}
\tablelasttail{
\bottomrule
}

\begin{supertabular}{@{\extracolsep{\fill}}l | rrr |rrrrrrr@{}}\label{tab:dimacs-data}%
1-FullIns\_3 & 30 & 100 & 0.23 & \textbf{0.1} & \textbf{0.1} & 0.3 & \textbf{0.1} & \textbf{0.1} & \textbf{0.1} & \textbf{0.1} \\
1-FullIns\_4 & 93 & 593 & 0.14 & \textbf{0.1} & \textbf{0.1} & 0.4 & \textbf{0.1} & \textbf{0.1} & - & \textbf{0.1} \\
1-FullIns\_5 & 282 & 3247 & 0.08 & \textbf{0.1} & 0.5 & 1.0 & \textbf{0.1} & 0.2 & - & \textbf{0.1} \\
1-Insertions\_4 & 67 & 232 & 0.10 & 158.1 & - & 1.2 & \textbf{0.4} & 1.5 & - & - \\
1-Insertions\_5 & 202 & 1227 & 0.06 & - & - & - & - & - & - & - \\
1-Insertions\_6 & 607 & 6337 & 0.03 & - & - & - & - & - & - & - \\
2-FullIns\_3 & 52 & 201 & 0.15 & \textbf{0.1} & \textbf{0.1} & 0.2 & \textbf{0.1} & \textbf{0.1} & \textbf{0.1} & \textbf{0.1} \\
2-FullIns\_4 & 212 & 1621 & 0.07 & \textbf{0.1} & 0.2 & 0.5 & \textbf{0.1} & \textbf{0.1} & - & \textbf{0.1} \\
2-FullIns\_5 & 852 & 12201 & 0.03 & 0.8 & 186.3 & 4.1 & \textbf{0.2} & 2.3 & - & 1.6 \\
2-Insertions\_3 & 37 & 72 & 0.11 & 0.2 & 0.4 & 0.2 & \textbf{0.1} & \textbf{0.1} & 273.5 & \textbf{0.1} \\
2-Insertions\_4 & 149 & 541 & 0.05 & - & - & - & - & - & - & - \\
2-Insertions\_5 & 597 & 3936 & 0.02 & - & - & - & - & - & - & - \\
3-FullIns\_3 & 80 & 346 & 0.11 & \textbf{0.1} & \textbf{0.1} & 0.2 & \textbf{0.1} & \textbf{0.1} & \textbf{0.1} & \textbf{0.1} \\
3-FullIns\_4 & 405 & 3524 & 0.04 & \textbf{0.1} & 0.5 & 1.1 & \textbf{0.1} & 0.2 & - & \textbf{0.1} \\
3-FullIns\_5 & 2030 & 33751 & 0.02 & \textbf{5.4} & 14.8 & 23.2 & \textbf{5.4} & 32.2 & - & - \\
3-Insertions\_3 & 56 & 110 & 0.07 & 1.0 & 7.0 & 0.2 & \textbf{0.1} & \textbf{0.1} & - & 2.3 \\
3-Insertions\_4 & 281 & 1046 & 0.03 & - & - & - & - & - & - & - \\
3-Insertions\_5 & 1406 & 9695 & 0.01 & - & - & - & - & - & - & - \\
4-FullIns\_3 & 114 & 541 & 0.08 & \textbf{0.1} & \textbf{0.1} & 0.3 & \textbf{0.1} & \textbf{0.1} & \textbf{0.1} & \textbf{0.1} \\
4-FullIns\_4 & 690 & 6650 & 0.03 & \textbf{0.1} & 1.9 & 2.3 & \textbf{0.1} & 0.8 & - & \textbf{0.1} \\
4-FullIns\_5 & 4146 & 77305 & 0.01 & \textbf{41.1} & - & 132.4 & 68.6 & 687.2 & - & - \\
4-Insertions\_3 & 79 & 156 & 0.05 & 11.3 & 2300 & 0.3 & 0.3 & \textbf{0.2} & - & 2145 \\
4-Insertions\_4 & 475 & 1795 & 0.02 & - & - & - & - & - & - & - \\
5-FullIns\_3 & 154 & 792 & 0.07 & \textbf{0.1} & \textbf{0.1} & 0.3 & \textbf{0.1} & \textbf{0.1} & \textbf{0.1} & \textbf{0.1} \\
5-FullIns\_4 & 1085 & 11395 & 0.02 & \textbf{0.2} & 3.8 & 5.9 & \textbf{0.2} & 5.9 & - & 0.3 \\
abb313GPIA & 1557 & 53356 & 0.04 & 305.9 & - & 24.9 & \textbf{0.6} & - & - & - \\
anna & 138 & 493 & 0.05 & \textbf{0.1} & \textbf{0.1} & 0.2 & \textbf{0.1} & \textbf{0.1} & \textbf{0.1} & \textbf{0.1} \\
ash331GPIA & 662 & 4181 & 0.02 & 15.0 & 2.5 & 1.8 & 2.0 & \textbf{0.1} & - & 0.2 \\
ash608GPIA & 1216 & 7844 & 0.01 & 128.1 & 27.3 & 5.0 & \textbf{0.2} & 0.3 & - & - \\
ash958GPIA & 1916 & 12506 & 0.01 & 1156 & 122.2 & 12.4 & \textbf{0.3} & 0.7 & - & 4.3 \\
C2000.5 & 2000 & 999836 & 0.50 & - & - & - & - & - & - & - \\
C2000.9 & 2000 & 1799532 & 0.90 & - & - & - & - & - & - & - \\
C4000.5 & 4000 & 4000268 & 0.50 & - & - & - & - & - & - & - \\
david & 87 & 406 & 0.11 & \textbf{0.1} & \textbf{0.1} & 0.2 & \textbf{0.1} & \textbf{0.1} & \textbf{0.1} & \textbf{0.1} \\
DSJC1000.1 & 1000 & 49629 & 0.10 & - & - & - & - & - & - & - \\
DSJC1000.5 & 1000 & 249826 & 0.50 & - & - & - & - & - & - & - \\
DSJC1000.9 & 1000 & 449449 & 0.90 & - & - & - & - & - & - & - \\
DSJC125.1 & 125 & 736 & 0.09 & 2.4 & 0.8 & 0.3 & 1.7 & \textbf{0.1} & 1713 & \textbf{0.1} \\
DSJC125.5 & 125 & 3891 & 0.50 & - & - & - & - & - & - & - \\
DSJC125.9 & 125 & 6961 & 0.90 & 508.1 & - & - & - & - & \textbf{7.0} & - \\
DSJC250.1 & 250 & 3218 & 0.10 & - & - & - & - & - & - & - \\
DSJC250.5 & 250 & 15668 & 0.50 & - & - & - & - & - & - & - \\
DSJC250.9 & 250 & 27897 & 0.90 & nan & - & - & - & - & - & - \\
DSJC500.1 & 500 & 12458 & 0.10 & - & - & - & - & - & - & - \\
DSJC500.5 & 500 & 62624 & 0.50 & - & - & - & - & - & - & - \\
DSJC500.9 & 500 & 112437 & 0.90 & - & - & - & - & - & - & - \\
DSJR500.1 & 500 & 3555 & 0.03 & \textbf{0.1} & \textbf{0.1} & 1.5 & \textbf{0.1} & \textbf{0.1} & 270.3 & \textbf{0.1} \\
DSJR500.1c & 500 & 121275 & 0.97 & \textbf{3.1} & - & - & \textbf{3.1} & 36.9 & 666.7 & - \\
DSJR500.5 & 500 & 58862 & 0.47 & 79.0 & - & - & - & \textbf{31.0} & 3167 & - \\
flat1000\_50\_0 & 1000 & 245000 & 0.49 & - & - & - & - & - & - & - \\
flat1000\_60\_0 & 1000 & 245830 & 0.49 & - & - & - & - & - & - & - \\
flat1000\_76\_0 & 1000 & 246708 & 0.49 & - & - & - & - & - & - & - \\
flat300\_20\_0 & 300 & 21375 & 0.48 & - & - & - & - & - & \textbf{629.1} & - \\
flat300\_26\_0 & 300 & 21633 & 0.48 & - & - & - & - & - & \textbf{1029} & - \\
flat300\_28\_0 & 300 & 21695 & 0.48 & - & - & - & - & - & - & - \\
fpsol2.i.1 & 496 & 11654 & 0.09 & \textbf{0.1} & 2.5 & 16.2 & \textbf{0.1} & \textbf{0.1} & 0.4 & \textbf{0.1} \\
fpsol2.i.2 & 451 & 8691 & 0.09 & \textbf{0.1} & 1.0 & 11.7 & \textbf{0.1} & \textbf{0.1} & 0.4 & \textbf{0.1} \\
fpsol2.i.3 & 425 & 8688 & 0.10 & \textbf{0.1} & 1.0 & 13.2 & \textbf{0.1} & \textbf{0.1} & 0.3 & \textbf{0.1} \\
games120 & 120 & 638 & 0.09 & \textbf{0.1} & \textbf{0.1} & 0.3 & \textbf{0.1} & \textbf{0.1} & \textbf{0.1} & \textbf{0.1} \\
homer & 561 & 1629 & 0.01 & \textbf{0.1} & \textbf{0.1} & 0.9 & \textbf{0.1} & \textbf{0.1} & \textbf{0.1} & \textbf{0.1} \\
huck & 74 & 301 & 0.11 & \textbf{0.1} & \textbf{0.1} & 0.3 & \textbf{0.1} & \textbf{0.1} & \textbf{0.1} & \textbf{0.1} \\
inithx.i.1 & 864 & 18707 & 0.05 & \textbf{0.1} & 3.5 & 21.4 & \textbf{0.1} & 0.2 & 1.2 & \textbf{0.1} \\
inithx.i.2 & 645 & 13979 & 0.07 & \textbf{0.1} & 2.0 & 18.7 & \textbf{0.1} & \textbf{0.1} & 0.3 & \textbf{0.1} \\
inithx.i.3 & 621 & 13969 & 0.07 & \textbf{0.1} & 1.9 & 19.3 & \textbf{0.1} & \textbf{0.1} & 0.3 & \textbf{0.1} \\
jean & 80 & 254 & 0.08 & \textbf{0.1} & \textbf{0.1} & 0.3 & \textbf{0.1} & \textbf{0.1} & \textbf{0.1} & \textbf{0.1} \\
latin\_square\_10 & 900 & 307350 & 0.76 & - & - & - & - & - & - & - \\
le450\_15a & 450 & 8168 & 0.08 & 21.8 & 13.2 & 4.1 & 10.1 & \textbf{0.2} & - & - \\
le450\_15b & 450 & 8169 & 0.08 & 10.6 & 28.1 & 4.0 & \textbf{0.2} & \textbf{0.2} & - & - \\
le450\_15c & 450 & 16680 & 0.17 & 696.1 & - & - & - & \textbf{58.7} & - & - \\
le450\_15d & 450 & 16750 & 0.17 & 289.8 & - & - & - & \textbf{45.0} & - & - \\
le450\_25a & 450 & 8260 & 0.08 & \textbf{0.1} & 0.5 & 4.9 & \textbf{0.1} & \textbf{0.1} & 1.4 & \textbf{0.1} \\
le450\_25b & 450 & 8263 & 0.08 & \textbf{0.1} & 0.3 & 4.9 & \textbf{0.1} & \textbf{0.1} & 2.2 & \textbf{0.1} \\
le450\_25c & 450 & 17343 & 0.17 & - & - & - & - & - & - & - \\
le450\_25d & 450 & 17425 & 0.17 & - & - & - & - & - & - & - \\
le450\_5a & 450 & 5714 & 0.06 & 6.9 & 30.3 & 2.0 & \textbf{0.1} & \textbf{0.1} & - & - \\
le450\_5b & 450 & 5734 & 0.06 & 5.3 & 34.5 & 1.9 & \textbf{0.1} & \textbf{0.1} & - & - \\
le450\_5c & 450 & 9803 & 0.10 & 2.9 & 5.6 & 3.0 & \textbf{0.1} & \textbf{0.1} & - & \textbf{0.1} \\
le450\_5d & 450 & 9757 & 0.10 & 2.3 & 5.1 & 3.0 & \textbf{0.1} & \textbf{0.1} & 3330 & 11.2 \\
miles1000 & 128 & 3216 & 0.40 & \textbf{0.1} & 0.5 & 5.8 & \textbf{0.1} & \textbf{0.1} & \textbf{0.1} & \textbf{0.1} \\
miles1500 & 128 & 5198 & 0.64 & \textbf{0.1} & 1.9 & 21.9 & \textbf{0.1} & \textbf{0.1} & \textbf{0.1} & \textbf{0.1} \\
miles250 & 128 & 387 & 0.05 & \textbf{0.1} & \textbf{0.1} & 0.3 & \textbf{0.1} & \textbf{0.1} & \textbf{0.1} & \textbf{0.1} \\
miles500 & 128 & 1170 & 0.14 & \textbf{0.1} & \textbf{0.1} & 0.4 & \textbf{0.1} & \textbf{0.1} & \textbf{0.1} & \textbf{0.1} \\
miles750 & 128 & 2113 & 0.26 & \textbf{0.1} & 0.2 & 2.3 & \textbf{0.1} & \textbf{0.1} & \textbf{0.1} & \textbf{0.1} \\
mug100\_1 & 100 & 166 & 0.03 & 0.6 & \textbf{0.1} & 0.3 & 0.6 & \textbf{0.1} & 0.6 & - \\
mug100\_25 & 100 & 166 & 0.03 & 0.6 & \textbf{0.1} & 0.3 & 0.5 & \textbf{0.1} & 0.5 & - \\
mug88\_1 & 88 & 146 & 0.04 & 0.4 & \textbf{0.1} & 0.3 & 0.3 & \textbf{0.1} & 0.3 & 371.7 \\
mug88\_25 & 88 & 146 & 0.04 & 0.4 & \textbf{0.1} & 0.3 & 0.4 & \textbf{0.1} & 0.4 & 241.7 \\
mulsol.i.1 & 197 & 3925 & 0.20 & \textbf{0.1} & 0.8 & 5.0 & \textbf{0.1} & \textbf{0.1} & \textbf{0.1} & \textbf{0.1} \\
mulsol.i.2 & 188 & 3885 & 0.22 & \textbf{0.1} & 0.4 & 5.4 & \textbf{0.1} & \textbf{0.1} & \textbf{0.1} & \textbf{0.1} \\
mulsol.i.3 & 184 & 3916 & 0.23 & \textbf{0.1} & 0.4 & 5.6 & \textbf{0.1} & \textbf{0.1} & \textbf{0.1} & \textbf{0.1} \\
mulsol.i.4 & 185 & 3946 & 0.23 & \textbf{0.1} & 0.4 & 5.7 & \textbf{0.1} & \textbf{0.1} & \textbf{0.1} & \textbf{0.1} \\
mulsol.i.5 & 186 & 3973 & 0.23 & \textbf{0.1} & 0.4 & 5.7 & \textbf{0.1} & \textbf{0.1} & \textbf{0.1} & \textbf{0.1} \\
myciel3 & 11 & 20 & 0.36 & \textbf{0.1} & \textbf{0.1} & 0.3 & \textbf{0.1} & \textbf{0.1} & \textbf{0.1} & \textbf{0.1} \\
myciel4 & 23 & 71 & 0.28 & \textbf{0.1} & \textbf{0.1} & 0.3 & \textbf{0.1} & \textbf{0.1} & 5.4 & \textbf{0.1} \\
myciel5 & 47 & 236 & 0.22 & \textbf{0.1} & \textbf{0.1} & 0.7 & \textbf{0.1} & 0.5 & - & 1.2 \\
myciel6 & 95 & 755 & 0.17 & \textbf{0.1} & \textbf{0.1} & 1981 & \textbf{0.1} & 932.6 & - & - \\
myciel7 & 191 & 2360 & 0.13 & \textbf{0.1} & 0.5 & - & \textbf{0.1} & - & - & - \\
qg.order100 & 10000 & 990000 & 0.02 & - & - & - & \textbf{1942} & - & - & - \\
qg.order30 & 900 & 26100 & 0.06 & 138.0 & \textbf{0.2} & 21.9 & 0.3 & 3.5 & - & 0.6 \\
qg.order40 & 1600 & 62400 & 0.05 & 1935 & 53.7 & 80.0 & \textbf{0.7} & 72.0 & - & - \\
qg.order60 & 3600 & 212400 & 0.03 & 1436 & 1710 & 2137 & \textbf{8.4} & 207.7 & - & - \\
queen10\_10 & 100 & 1470 & 0.30 & 183.4 & - & 371.3 & - & - & \textbf{170.3} & - \\
queen11\_11 & 121 & 1980 & 0.27 & \textbf{112.1} & - & 346.4 & - & - & - & - \\
queen12\_12 & 144 & 2596 & 0.25 & - & - & - & - & - & - & - \\
queen13\_13 & 169 & 3328 & 0.23 & - & - & - & - & - & - & - \\
queen14\_14 & 196 & 4186 & 0.22 & - & - & - & - & - & - & - \\
queen15\_15 & 225 & 5180 & 0.21 & - & - & - & - & - & - & - \\
queen16\_16 & 256 & 6320 & 0.19 & - & - & - & - & - & - & - \\
queen5\_5 & 25 & 160 & 0.53 & \textbf{0.1} & \textbf{0.1} & 0.4 & \textbf{0.1} & \textbf{0.1} & \textbf{0.1} & \textbf{0.1} \\
queen6\_6 & 36 & 290 & 0.46 & \textbf{0.1} & 0.3 & 0.5 & \textbf{0.1} & \textbf{0.1} & 0.3 & \textbf{0.1} \\
queen7\_7 & 49 & 476 & 0.40 & \textbf{0.1} & \textbf{0.1} & 0.6 & \textbf{0.1} & \textbf{0.1} & 0.6 & \textbf{0.1} \\
queen8\_12 & 96 & 1368 & 0.30 & 0.3 & \textbf{0.1} & 1.0 & \textbf{0.1} & \textbf{0.1} & 7.7 & \textbf{0.1} \\
queen8\_8 & 64 & 728 & 0.36 & 1.3 & 5.8 & 3.4 & \textbf{0.4} & 12.7 & 5.2 & 1.0 \\
queen9\_9 & 81 & 1056 & 0.33 & 25.8 & 160.3 & 11.7 & 381.9 & - & \textbf{11.1} & 588.8 \\
r1000.1 & 1000 & 14378 & 0.03 & \textbf{0.1} & 0.5 & 9.4 & \textbf{0.1} & 0.2 & 1.0 & 0.8 \\
r1000.1c & 1000 & 485090 & 0.97 & - & - & - & - & - & - & - \\
r1000.5 & 1000 & 238267 & 0.48 & - & - & - & \textbf{847.5} & 1870 & - & - \\
r125.1 & 125 & 209 & 0.03 & \textbf{0.1} & \textbf{0.1} & 0.3 & \textbf{0.1} & \textbf{0.1} & \textbf{0.1} & \textbf{0.1} \\
r125.1c & 125 & 7501 & 0.97 & \textbf{0.1} & 3.5 & 37.9 & \textbf{0.1} & \textbf{0.1} & \textbf{0.1} & \textbf{0.1} \\
r125.5 & 125 & 3838 & 0.50 & 4.3 & 2.7 & 8.9 & 0.3 & \textbf{0.1} & 11.6 & \textbf{0.1} \\
r250.1 & 250 & 867 & 0.03 & \textbf{0.1} & \textbf{0.1} & 0.5 & \textbf{0.1} & \textbf{0.1} & \textbf{0.1} & \textbf{0.1} \\
r250.1c & 250 & 30227 & 0.97 & \textbf{0.1} & 40.3 & 103.3 & \textbf{0.1} & 0.3 & 30.7 & 1.0 \\
r250.5 & 250 & 14849 & 0.48 & 94.5 & 36.0 & 71.1 & 3.0 & \textbf{2.6} & 185.9 & 199.6 \\
school1 & 385 & 19095 & 0.26 & \textbf{0.3} & 3.1 & 18.2 & \textbf{0.3} & 0.6 & 968.9 & 5.7 \\
school1\_nsh & 352 & 14612 & 0.24 & \textbf{0.1} & 4.8 & 11.6 & \textbf{0.1} & \textbf{0.1} & 1840 & 6.0 \\
wap01a & 2368 & 110871 & 0.04 & - & - & 1778 & - & \textbf{124.1} & - & - \\
wap02a & 2464 & 111742 & 0.04 & - & - & 1054 & - & \textbf{144.4} & - & - \\
wap03a & 4730 & 286722 & 0.03 & - & - & - & - & - & - & - \\
wap04a & 5231 & 294902 & 0.02 & - & - & - & - & - & - & - \\
wap05a & 905 & 43081 & 0.11 & \textbf{0.1} & 5.6 & 61.2 & \textbf{0.1} & 0.7 & 5.2 & 0.9 \\
wap06a & 947 & 43571 & 0.10 & 413.0 & - & 114.0 & 575.5 & \textbf{31.6} & - & - \\
wap07a & 1809 & 103368 & 0.06 & - & - & \textbf{-} & - & - & - & - \\
wap08a & 1870 & 104176 & 0.06 & - & - & \textbf{504.9} & 2345 & - & - & - \\
will199GPIA & 701 & 6772 & 0.03 & 0.3 & 1.2 & 3.2 & 0.9 & \textbf{0.2} & 1.5 & 0.3 \\
zeroin.i.1 & 211 & 4100 & 0.19 & \textbf{0.1} & 0.9 & 5.5 & \textbf{0.1} & \textbf{0.1} & \textbf{0.1} & \textbf{0.1} \\
zeroin.i.2 & 211 & 3541 & 0.16 & \textbf{0.1} & 0.4 & 4.3 & \textbf{0.1} & \textbf{0.1} & \textbf{0.1} & \textbf{0.1} \\
zeroin.i.3 & 206 & 3540 & 0.17 & \textbf{0.1} & 0.4 & 4.5 & \textbf{0.1} & \textbf{0.1} & \textbf{0.1} & \textbf{0.1} \\
\end{supertabular}
\end{center}

}

\end{document}